\documentclass[aps,prl,twocolumn,showpacs,preprintnumbers,amsmath,amssymb]{revtex4}
\usepackage{bm}
\usepackage{epsfig,amsopn}
\usepackage{graphicx}
\usepackage{amsmath,amssymb}
\usepackage{natbib}
\renewcommand{\section}[1]{{\par\it #1.---}}
\def\be{\begin{equation}}
\def\ee{\end{equation}}
\def\bea{\begin{eqnarray}}
\def\eea{\end{eqnarray}}
\def\la{\langle}
\def\ra{\rangle}
\def\om{\omega}
\def\nn{\nonumber}

\def\f{\frac}
\def\g{\gamma}
\def\al{\alpha}
\def\etal{et al.}
\def\n{\eta}
\def\l{\lambda}

\begin{document}
\title{Effect of phonon-phonon interactions on localization}
\author{Abhishek Dhar}
\affiliation{Raman Research Institute, Bangalore 560080, India}
\author{J.L. Lebowitz}
\affiliation{Department of Mathematics and Physics, 
    Rutgers University, Piscataway, NJ 08854}
\date{\today}
\begin{abstract}
We study the heat current $J$ in a classical one-dimensional disordered chain  
with 
onsite pinning and with ends connected to stochastic thermal reservoirs at 
different temperatures. In the absence of anharmonicity all modes are 
localized and there is a gap in the spectrum.  Consequently  $J$ 
decays exponentially with system size $N$. 
Using simulations we find that even a small amount of anharmonicity 
leads to a $J\sim 1/N$ dependence, implying diffusive transport 
of energy. 
\end{abstract}

\maketitle
The effect of interactions between electrons or phonons on localization 
caused by disorder is a subject of both theoretical 
\cite{fleishman80,LLA78,GS88,NGD91,GMP05,basko06,vadim07,pikovsky07,kopidakis07,song00,SBZ91,payton67} 
and experimental interest \cite{expt}. As is well known localization of
eigenfunctions or  of normal modes 
strongly affects transport in materials containing random impurities. 
Localization, first discovered in electronic systems by Anderson
\cite{anderson58},  has its strongest effect in one dimensions where 
any finite disorder makes all eigenstates localized 
\cite{mott61} and one has an insulator. The presence of
inelastic scattering, such as  is caused by electron-phonon interactions, 
leads to hopping of electrons between localized states and
gives rise to a finite conductivity. The question as to whether
electron-electron interactions  lead to a similar effect
\cite{LLA78,NGD91} has attracted much attention recently
 but is still not fully understood \cite{GMP05,basko06,vadim07,pikovsky07,kopidakis07}.  

In this paper we address the same question in the context of heat
conduction by phonons
and consider the effect that phonon-phonon interactions have on localization.  
In particular we investigate the effect of anharmonicities on the steady state 
transport of heat through a chain of oscillators with random masses.
We focus on the case where the masses are subjected to an external pinning
potential, in addition to nearest neighbor interactions. Pinning greatly 
enhances the  difference between heat transport in a random chain 
with and without anharmonicity 
and thus is a good testing ground for the effect of anharmonicity on 
localization.  We also discuss the unpinned case and comment on 
results from some earlier studies.

The Hamiltonian of our system has the form
\bea
H &=& \sum_{l=1,N} [ \f{p_l^2}{2 m_l} + k_o \f{x_l^2}{2} + 
\lambda \f{x_l^4}{4} ]\nn \\
&+&  \sum_{l=1,N+1}
[ ~ k \f{(x_l-x_{l-1})^2}{2} +\nu \f{(x_l-x_{l-1})^4}{4} ~] \label{ham}
\eea
where $\{x_l,p_l\}$ denote the position and momenta of the particles and 
we set $x_0=x_{N+1}=0$.
The masses $\{m_l\}$ are chosen independently from some 
distribution $p(m)$, {\emph{e.g.}} one uniform in the interval
$(m-\Delta,m+\Delta)$. 
The chain is connected at its ends to two heat baths at temperatures 
$T_L$ and $T_R$ 
respectively. The baths will be modeled by Ornstein-Uhlenbeck (Langevin 
white noise) reservoirs.
The equations of motion of the chain are then given by:
\bea
\label{eq: 2}
m_l \ddot{x}_l&=&- k_o x_l - \l x_l^3-k (2 x_l-x_{l-1}-x_{l+1}) \nn \\
&-& \nu [ (x_l-x_{l-1})^3 + (x_l-x_{l+1})^3 ]-\g_l \dot{x}_l 
+\n_l~,  
\eea
with $\n_l=\n_L \delta_{l,1}+\n_R \delta_{l,N},~\g_l=\g (\delta_{l,1}+
\delta_{l,N})$, 
 and where the Gaussian noise terms satisfy the  fluctuation dissipation 
relations $\la \eta_L(t) \eta_L(t') \ra = 2 \g k_B T_L \delta(t-t')$,
$\la \eta_R(t) \eta_R(t') \ra = 2 \g k_B T_R \delta(t-t')$, $k_B$
being Boltzmann's constant. It is known that 
this system has a unique stationary state \cite{EPR99}.

Here
we investigate the $N$ dependence of the heat current and the 
temperature profile, in the nonequilibrium stationary state (NESS) of this 
system, when $T_L > T_R$. The heat current from left to right is given by 
$ \la J_N \ra = \sum_l \la f_{l,l-1} \dot{x}_l \ra/(N-1)$ where $f_{l,l-1}$ 
is the force exerted 
by the $(l-1)$th particle on the $l$th particle and $\la ...\ra$ denotes an 
average over the NESS. It follows from stationarity that each term in the sum,
$l=2,...,N$ is equal to $\la J_N \ra$.  

Note that Eq.(\ref{eq: 2}) is invariant under the transformation $T_{L,R} \to s T_{L,R}$,
$\{ x_l\} \to \{ s^{1/2} x_l \}$ and $(\l,\nu) \to  (\l,\nu)/s$.  This implies
the scaling relation $\la J_N (sT_L,sT_R,\l,\nu)\ra = s \la
J_N (T_L,T_R,s\l,s \nu)\ra$. Thus the effect of changing nonlinearity
could be equivalently studied by changing temperatures.

In the  harmonic case, $\lambda=\nu=0$, the
scaling gives $\la
J_N\ra$ proportional to $(T_L - T_R)$. 
In this case the quadratic Hamiltonian can be written in the form 
$H_0=({1}/{2}) [ P^T M^{-1} P + X^T \Phi X]$ using matrix notation for the mass
and force matrices.
The stationary heat current, for any given realization of disorder
is then given by the following expression \cite{casher71,dhar01}:
\bea
\la J_N \ra&=&\f{k_B(T_L-T_R)}{\pi} \int_{-\infty}^\infty d \om {\cal{T}}_N(\om)~,
\label{genexp} 
\\   
{\rm where} &&~~{\cal{T}}_N(\om) = \g^2 \om^2 |G_{1N}|^2 ~,  \nn 
\eea
with the matrix $G=[-\om^2 M + \Phi - \Sigma]^{-1}$  and 
$\Sigma_{lm}=i \gamma \om \delta_{lm} [\delta_{l1}+\delta_{lN}]$.

The NESS in the harmonic chain without randomness, {\emph{ i.e}} $m_l=m$ can
be solved exactly. It gives $\la J_N  \ra \to c  k_B
(T_L-T_R)$, with 
$ c >0$, for $N\to \infty$ \cite{rieder67,nakazawa70}~, {\emph{i.e.}}
the conductivity grows linearly with $N$.  

The nature of $\la J_N \ra$ for the random harmonic chain without pinning 
was analyzed in much detail 
in \cite{casher71,rubin71,connor74,dhar01}. In \cite{casher71} it was proved 
that, for an arbitrary nontrivial random mass distribution, 
$\la J_N\ra \to 0$, with $\la J_N \ra \geq N^{-3/2}$.  The results in \cite{casher71,rubin71,
connor74,dhar01} are based on the use of the  Furstenberg theorem 
\cite{furstenberg63} first introduced 
into this problem by Matsuda and Ishii \cite{matsuda70}. (In fact, Casher and 
Lebowitz use the $\la J_N \ra \to 0$ as $N \to \infty$ result to prove the 
absence of an absolutely continuous part of the spectrum for the random 
semi-infinite chain.) The fact that the random $1$D system has no extended 
states is of course well-known. The mathematical proof given in  
 \cite{goldshtein73}
shows that such a chain has pure point spectrum, {\emph{i.e.}} all
the eigenfunctions are square integrable. The flux $\la J_N \ra$ in this
 chain 
is  carried entirely by the long wave-length modes with frequencies
$\om \stackrel{<}{\sim} N^{-1/2}$ which ``do not see'' the randomness. More 
detailed results 
about this case can be found in
\cite{casher71,rubin71,connor74,dhar01}.  In particular, based on
numerical evidence \cite{dhar01,likhachev}
we believe that indeed $\la J_N \ra 
\sim N^{-3/2}$.   We note however that $\la J_N\ra$ in the unpinned 
harmonic case with disorder depends on the 
particular type of heat bath \cite{dhar01}.  For 
the Rubin model \cite{rubin71,verheggen79}, where  the
baths are semi-infinite ordered 
harmonic chains in equilibrium at temperatures $T_L (T_R)$ at the left
(right) ends of the system and with no pinning anywhere, one gets $\la
J_N\ra\sim N^{- 1/2}$.

Considering now the random mass harmonic pinned case, $k_o >0$, one can 
show that  
$\la J_N \ra \sim e^{-c N},~ c>0 $.   
This follows from the fact that the spectrum of this chain now lies 
entirely in the interval $[(\f{k_o}{\overline{m}})^{1/2},
(\f{k_o+2 k}{\underline{m}})^{1/2}]$  where $\overline{m} 
(\underline{m})$ are the maximum (minimum) of the masses $\{ m_l\}$.
This means that there is a gap in the spectrum and Furstenberg's theorem
implies that ${\cal{T}}_N(\om) \stackrel{<}{\sim} 
e^{-N \delta (\om)}~, \delta (\om) > 0$ for all $\om$ in the allowed
frequency range. 
An elementary calculation shows that
${\cal{T}}_N (\om) \sim e^{-N a}~, a>0$ for $\om$ outside the spectrum 
(true for both 
the ordered and disordered case). Hence, in the disordered harmonic 
pinned case, we will have asymptotically for almost all realizations of
disorder 
\bea
-\f{1}{N} \ln \la J_N \ra \sim c = \ell^{-1}~,
\eea 
where we can interpret $\ell$ as the  largest localization length. 
A rough estimate of the value 
of $\ell$ can be obtained using known results for the unpinned case at small 
$\om$. Substituting in these results the smallest allowed frequency  
$\om_m^2=k_o/\overline{m}$ we get   
$\ell = 12 k m /(\om_m^2 \Delta^2)$  for masses chosen from a uniform
distribution between $[m-\Delta,m+\Delta]$. 
This can be compared with the  results from  a numerical 
evaluation  of the integral in Eq.~(\ref{genexp}) for the  case 
with $k_o=k=1~,m=1$ and $\Delta=0.2$.  
In Fig.~(1) ($\lambda=0$ data) we have plotted $N[\la J_N\ra]$, 
averaged  over $100$ disorder realizations, as a function of system
size. From the data we find an exponential decay  with $\ell \approx 200$
while our very rough  
estimate gives $\ell \approx 360$.

For the anharmonic chain there are no 
rigorous result about $\la J_N \ra$, even for the ordered case
but the  general expectation is that 
$\la J_N \ra \sim N^{\al-1}$ for the unpinned case. The actual value of $\al$ 
is a matter of some dispute with values ranging in the interval 
$\alpha =1/3-1/2$ \cite{NR02,LLP03,BBO06}. The most recent simulations
found $\alpha=1/3$  \cite{mai07}. 

A number of simulations for the disordered unpinned case with $ \nu > 0,
k_o=\lambda =0$ show that anharmonicity in general tends to destroy 
localization \cite{payton67}.  
In a recent work, 
Li {\emph{ et al} \cite{baowenli01} found a transition from $\al=0$ 
at small 
anharmonicity to $\alpha \sim 0.43$ at large anharmonicity.    
They did the simulations using a deterministic model for the reservoirs, 
namely Nose-Hoover thermostats. However this model of thermostats has been 
shown in \cite{dharb01} to be problematic for the harmonic case 
 and these problems persist for small anharmonicity as well \cite{dharc07}.
Preliminary simulations carried out  for the unpinned disordered 
anharmonic case,  with white noise 
Langevin heat baths, do not  find evidence  of a finite $\nu$  transition
\cite{dharWIP07}.  

We now consider the pinned case $k= k_o=1, \lambda >0, \nu=0$ when the 
system without randomness is much better understood. It is generally
agreed that, in the absence of disorder, models with onsite pinning and anharmonicity
show regular heat transport with $\la J_N \ra \sim 1/N$ 
\cite{casati84,aoki06}.  This is proven rigorously for the case when a
certain amount of stochasticity is added to the dynamics \cite{bonetto}.
This model is also closer in spirit to charge transport by hopping in
random media \cite{anderson58,mott61} .

In our simulations  with both  disorder and anharmonicity 
we  fixed the  values of $k,T_R,\la m
\ra,\gamma$. We then measure time and distance in units of
$(\la m \ra/k)^{1/2}$ and $(k_B T_R 
\gamma)^{1/2}/k$ respectively and set $k_o/k=1,~T_L/T_R=2$. The only
free parameters $\Delta$ and $\lambda$ are measured in units of $\la m
\ra$ and $k^3/(k_BT_R \gamma)$ respectively (in the simulations we set
$k=k_o=\la m \ra=\gamma=T_R=1$).   
In Fig.~(\ref{JvsN}) we plot  $N [<J_N>]$ as a function of $N$ 
for a fixed disorder strength $\Delta=0.2$ and different values of
anharmonicity  $\lambda=0.004-1.0$. 
As can be seen from our data, there is a 
dramatic {\it{increase}} in the heat current on introduction of a small amount
of anharmonicity and the system-size dependence goes from exponential decay to
a $1/N$ dependence implying diffusive transport. For smaller $\lambda$
the diffusive regime sets in at larger length scales. Similar results
are obtained for the case with $\lambda=0,~\nu >0$ and are shown in the
inset of Fig.~(\ref{JvsN}).

The simulations were done by the velocity Verlet algorithm adapted for
Langevin dynamics \cite{AT87}.
Equilibration times ranged from $10^8-2\times 10^8$ time steps of
step-size $0.005-0.01$ and steady state averages were taken  
over another $10^8-8 \times 10^8$ time steps. 
Equilibration times increase rapidly with decreasing $\lambda$ and
with increasing $N$ and these cases required the longest runs.
The error bars correspond to 
sample to sample fluctuations (other errors are smaller) 
and the number of samples varied from $10$ for 
the small sizes to $3$ for the largest ($N=4096$). We find that the
sample-sample  
fluctuations become smaller with increasing $N$. 
A measure of the relative 
strengths of anharmonicity and disorder is obtained by looking at the 
ratio of the energy scales  $E_a= \l \la x^4 \ra/4$ and $E_d=
T \Delta/m $. For our parameters we estimate 
$\epsilon= E_a/E_d \approx 0.3 \lambda T/\Delta$ and for
$T=(T_L+T_R)/2=1.5$ this gives  $\epsilon \approx 0.008$ for $\lambda =0.004$.

It follows from the scaling relation described earlier that 
$\la J_N \ra = (T_L-T_R) f(\lambda T_L, \l T_R)$. Hence the
conductivity, for $N \to \infty$ and $T_L \to T_R=T$, should depend only on $\lambda T$,
$\kappa = \kappa( \l T)$. For $T_L > T_R$ the temperature and hence
the conductivity  varies across the chain, we
therefore measure the effective thermal 
conductivity $\kappa_{eff}=N [<J_N>]$, evaluated at large $N$
(boundary temperature jumps are negligible). 
For small $\lambda$  we find 
$\kappa_{eff} \sim (\l T)^a$ with  $a< 1$. Assuming that this is also
the behaviour of $\kappa$ this implies, using the
fact that  $\kappa dT/dy=$constant, that the temperature profile is
given by $T(y)=[T_L^{a+1} (1-y)
  +T_R^{a+1} y ]^{1/(a+1)}$ where $y$ is the distance, scaled by $N$,
from the left end of the chain. 
We now look at the local temperature profiles  in the 
NESS obtained in the simulations. The local temperature at the $i$th
site is defined by $T_i=m_i \la \dot{x}_i^2 \ra$.
The temperature profiles for $\l=0.1$ for a single sample, with and
without disorder, are plotted in Fig.~(\ref{TempPr}).  The plots
(a),(b) correspond respectively to averaging over $8\times 10^8$ and
$32 \times 10^8$ time steps with $dt=0.005$ and show that the profiles
are reasonably converged.The noise amplitude does not decrease much
on increased averaging.  
We find that the temperature profile, for the disordered case, is
consistent with the predicted 
form with $a \sim 1/2 $ (see however below).

For the case without disorder we find a temperature profile
consistent with a $1/T^2$ (thus $a=-2$)  dependence for the conductivity.  The 
$\kappa(T)\sim1/T^2$ dependence has been
predicted from recent kinetic theory calculations 
and should apply for small 
nonlinearity \cite{aoki06}.  
The inset in Fig.~(\ref{TempPr}) shows temperature profiles for the
disordered chain with a  smaller  and a larger value of $\lambda$. 
We find that these profiles have some amount of structure, possibly
reflecting the local mass distribution, and vary appreciably from
sample to sample. 
\begin{figure}
\vspace{0.25cm}
\includegraphics[width=3in]{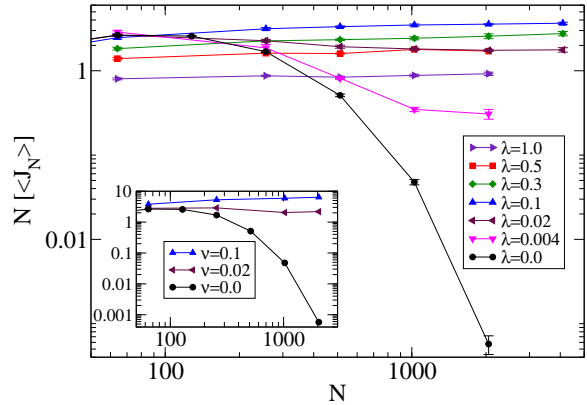}
\caption{Plot of the disordered averaged heat current $[\la J_N\ra ]$
  multiplied by $N$ as a function of 
$N$ for different values of $\lambda$. The inset shows results
  obtained for the case with interparticle anharmonicity.
}   
\label{JvsN}
\end{figure}


\begin{figure}
\vspace{0.25cm}
\includegraphics[width=3in]{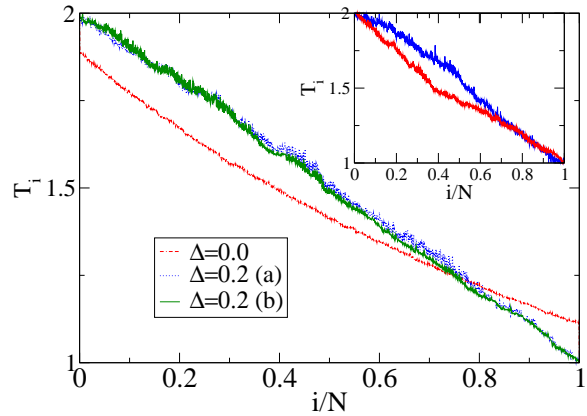}
\caption{The temperature profile for an anharmonic chain with 
$\lambda=0.1$ and $N=2048$ is shown for both the ordered and disordered cases. 
The inset shows the temperature profiles for a smaller value of
anharmoncity $\lambda=0.02$ (upper curve) and a larger value
$\lambda=0.5$. In all disordered cases $\Delta=0.2$.
}   
\label{TempPr}
\end{figure}

{\emph{Discussion}}: We have studied heat conduction in a 
disordered pinned anharmonic chain. We find that  introduction of a
small amount of 
 phonon-phonon interactions in the disordered harmonic chain  leads to
 diffusive energy transfer, {\emph {i.e.}} the insulating chain becomes a normal heat 
conductor. We do not find evidence of the existence of a finite
critical value of anharmonicity  required for this transition. 
For small values of anharmonicity 
it is necessary to go to larger system sizes to see the transition 
from insulating to diffusive. Hence a transition to a localized phase
at a very small value of anharmonicity is possible and would be
difficult to observe in simulations. Even assuming that no such
transition occurs, the limiting behavior of $\kappa (\lambda
T)$ for $(\lambda T)\to 0$ cannot be obtained from our data. A. Pal
and D. Huse \cite{pal} have investigated this type of question for a chain
of classical spins.  Their system corresponds very roughly to 
the case where $k= \lambda = 0$ and so the interaction between
particles on different sites
only comes via $\nu$: they find a 
diffusivity which behaves like 
$e^{-c/\nu}$.  We remark that for the ordered harmonic chain letting
$k\to 0$ gives $\la J_N\ra\sim k$ for the unpinned case
and $\la J_N\ra\sim k^2$ for the pinned case.

We note that in the ordered
chain nonlinearity leads to scattering of otherwise freely moving
phonons thereby changing the system from a super-heat conductor to a
normal one. By contrast, 
in the disordered chain the nonlinearity appears 
to produce effective  extended states which  lead to 
diffusive transfer of energy between local regions. How exactly this
occurs is not clear. 
 
The diffusive transport
observed in our disordered open system seems to differ from
recent numerical studies
of spreading of localized energy pulses in disordered nonlinear
lattices which suggest sub-diffusion \cite{pikovsky07} or even absence of
diffusion \cite{kopidakis07}. 
To understand this  
we have also performed some studies on our system without
heat baths. From these we find that  while a heat pulse introduced at
one end of a long chain does indeed {\emph {not}} propagate,  a periodic driving at one
end leads to significant energy transmission. 
This  suggests that  an isolated pulse  behaves differently from a
continuously driven or open system \cite{zhao06}.

Finally we note that experimental measurements of heat conduction in
one-dimensional systems are now becoming possible
\cite{angel98,schwab,wang07}. Interesting measurable effects which our
study suggests are the effect of temperature and system-size on heat
conduction in disordered wires. Both of these enhance the
effect of nonlinearity and lead to diffusive transport.

We thank M. Aizenmann, E. Lieb, J. Lukkarinen, V. Oganesyan,
T. Spencer, H. Spohn and especially D. Huse
for useful discussions. AD thanks K. Saito and B. S. Shastry for
discussions. Research 
supported in part by NSF grant 0442066 and AFOSR grant AF-FA9550-04.

\end{document}